\documentclass[twoside,a4paper,11pt]{torus2012}
\usepackage{graphicx}
\usepackage{hyperref}
\usepackage{movie15}
\begin{document}
\pagenumbering{arabic}
\pagestyle{myheadings}
\thispagestyle{empty}
{\flushleft\includegraphics[width=\textwidth,bb=58 650 590 680]{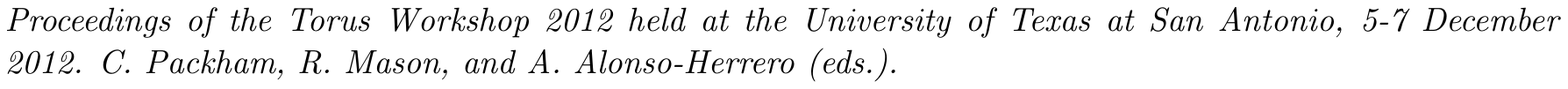}}
\vspace*{0.2cm}
\begin{flushleft}
{\bf {\LARGE
%
The largest mid-infrared atlas of active galactic nuclei at sub-arcsecond spatial scales
%
}\\
\vspace*{1cm}
%
Daniel Asmus$^{1}$,
Poshak Gandhi$^{2}$,
Sebastian F. H\"onig$^{3}$
and
Alain Smette$^{4}$
%
}\\
\vspace*{0.5cm}
%
$^{1}$
Max-Planck-Institut f\"ur Radioastronomie, 
           Auf dem H\"ugel 69, 53121 Bonn, Germany \\
$^{2}$
Institute of Space and Astronautical Science (ISAS), Japan Aerospace Exploration
Agency, 3-1-1 Yoshinodai, chuo-ku, Sagamihara, Kanagawa
252-5210, Japan\\
$^{3}$
UCSB Department of Physics, Broida Hall 93106-9530, Santa Barbara, CA, USA\\
$^{4}$
European  Southern Observatory, Casilla 19001, Santiago 19, Chile
%
\end{flushleft}
%
\markboth{
The largest mid-infrared atlas of active galactic nuclei at sub-arcsecond spatial scales
}{ 
%
Asmus et al.
%
}
\thispagestyle{empty}
\vspace*{0.4cm}
\begin{minipage}[l]{0.09\textwidth}
\ 
\end{minipage}
\begin{minipage}[r]{0.9\textwidth}
\vspace{1cm}
\section*{Abstract}{\small
%
We present the largest mid-infrared atlas of active galactic nuclei at sub-arcsecond spatial scales containing 249 objects. 
It comprises all ground-based HR MIR observations performed to date. This catalog
includes a large number of new observations. The photometry in multiple filters
allows for characterizing the properties of the dust emission for most objects.
Because of its size and characteristics, this sample is very well-suited for AGN
unification studies. In particular, we discuss the enlarged MIR--X-ray
correlation which extends over six orders of magnitude in luminosity and
potentially probes different physical mechanisms. Finally, tests for intrinsic
differences between the AGN types are presented and we discuss dependencies of
MIR--X-ray properties with respect to fundamental AGN parameters such as
accretion rate and the column density and covering factor of obscuring material.
%
\normalsize}
\end{minipage}
%
%
%

\section{Introduction \label{intro}}
Mid-infrared (MIR) observations of active galactic nuclei (AGN) enable the study
of the astrophysical dust in these objects. This dust plays a key role in our understanding of
the central accreting supermassive black hole and the surrounding star formation (SF). 
During the last decade, an increasing number of works have demonstrated the power and importance of high-angular resolution (HR) MIR observations in order to isolate the AGN from surrounding starbursts, e.g., \cite{Gorjian04, Gandhi09, Levenson09, RamosA11, Asmus11} (see also contribution by Ramos Almeida et al.)
However, owing to the comparably low sensitivity and higher complexity of ground-based MIR observations (contrary to low-angular resolution space-based with, e.g., \textit{Spitzer}), only relatively small samples have been observed and analyzed so far. 

Furthermore, while the HR enables us to resolve non-AGN emission regions in the nuclear regions of nearby galaxies, the AGN itself remains mostly unresolved even with 8\,meter class telescopes (apart from the outer narrow line region).
Many components of the AGN can possibly emit significant MIR emission, starting from the accretion disk surrounding the central supermassive black hole (SMBH), the perpendicularly emitted highly beamed jet outflow to the emission line clouds and the dusty obscuring structure ('torus') . 

Finally, MIR interferometric observations enable us to resolve the AGN components but due to the sensitivity limits, only a relatively small number of AGN can be studied (contributions by  Burtscher et al., Kishimoto et al., and Tristram et al.).

For these reasons, this project aims at understanding the MIR emitting structure in AGN by assembling HR MIR spectral energy distributions for a large representative sample of the local AGN.
This can then be used for multiwavelength studies as presented here, in particular the
MIR--X-ray relation. 
The full analysis of this dataset will be published in Asmus et al. (in prep,
a,b).

\section{Sample selection and observations}
High angular resolution is of uttermost importance for the study of the MIR properties of local AGN. 
Therefore, we have considered imaging observations of only the largest
single dish facilities in order to amass an AGN atlas of all ground-based  MIR observations ever taken. 
In particular, we concentrate on facilities having public archives:
Gemini/Michelle\cite{Glasse97},
 Gemini/T-ReCS\cite{Telesco98},
 Subaru/COMICS\cite{Kataza00},
 and VLT/VISIR\cite{Lagage04}.
 
The base sample of this study is the uniform BAT 9-month AGN sample consisting of 104 objects\cite{Winter09}.
Of those, we have observed a subsample of 80 sources with at least one of the above
instruments during the last years. 
Despite its selection method at hardest X-rays (14-195\,keV), the BAT AGN
sample under-represents the highest absorbed part of the AGN population, in
particular Compton-thick objects. 
Furthermore, the rather high flux limit leads to a cutoff of most
low-luminosity AGN, which represent the majority of the nearby AGN.
For these reasons, we have complemented the BAT AGN sample with all local AGN
that have imaging observations available in at least one of the four instruments
(COMICS, Michelle, T-ReCS and VISIR). 
In this work, we define 'local' as redshift $z \le 0.4$. 
The optical classifications are mainly obtained from \cite{Veron10}.
This selection leads to a total sample of 249 AGN with a median redshift of
0.016, the 'AGN MIR atlas'.
Note that this atlas is not complete in terms of volume or flux thresholds but simply contains 'everything that has been observed'. 
However, with respect to \cite{Veron10} the AGN MIR atlas contains more than one third of all
optically identified AGN with a redshift $<0.01$.
Furthermore, it is sufficiently large to allow construction of well-matched samples based upon various selection strategies in future investigations.

In the following of this work, we only distinguish between type~1 (Sy 1.0-1.5, and 1n), type~2 (Sy 1.8-2.0
and 1h, 1i) AGN and LINERs (low-ionization nuclear emission line regions).
With this definition, the majority of the objects are type~2.

In total, about 1000 N- and Q-band images have been analyzed, of which more than 600 are so far unpublished. 
The majority of images have been obtained with VISIR, followed by T-ReCS,
Michelle, and the least with COMICS.
All observations have been carried out in standard chopping and nodding mode.
Usually a flux standard star was observed either before or after the AGN within
two hours and is used for the flux calibration. 
The flux measurements are carried out in the same way as described in
\cite{Asmus11} using a custom Gaussian fitting photometry method.

\section{Results}
\begin{figure}[htp]
\center
\includegraphics[scale=0.79]{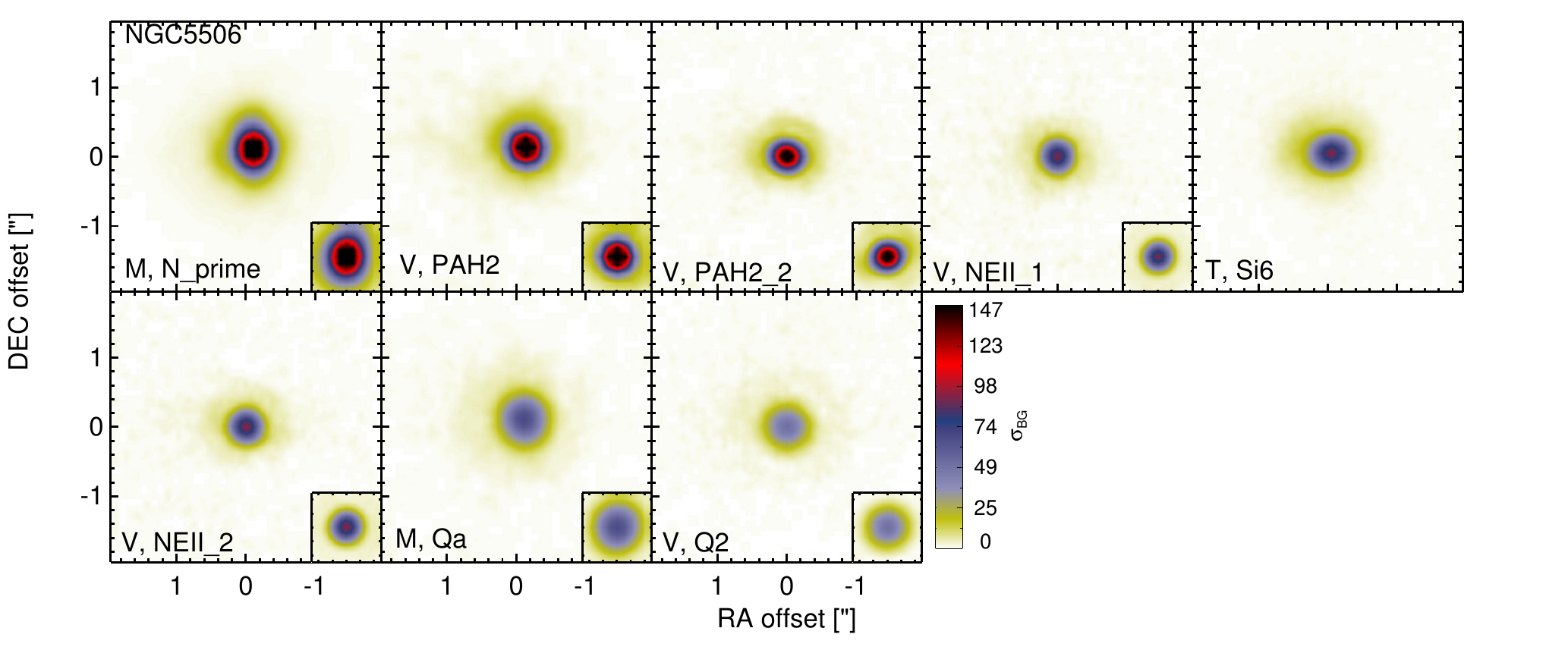} ~
\includegraphics[scale=0.9]{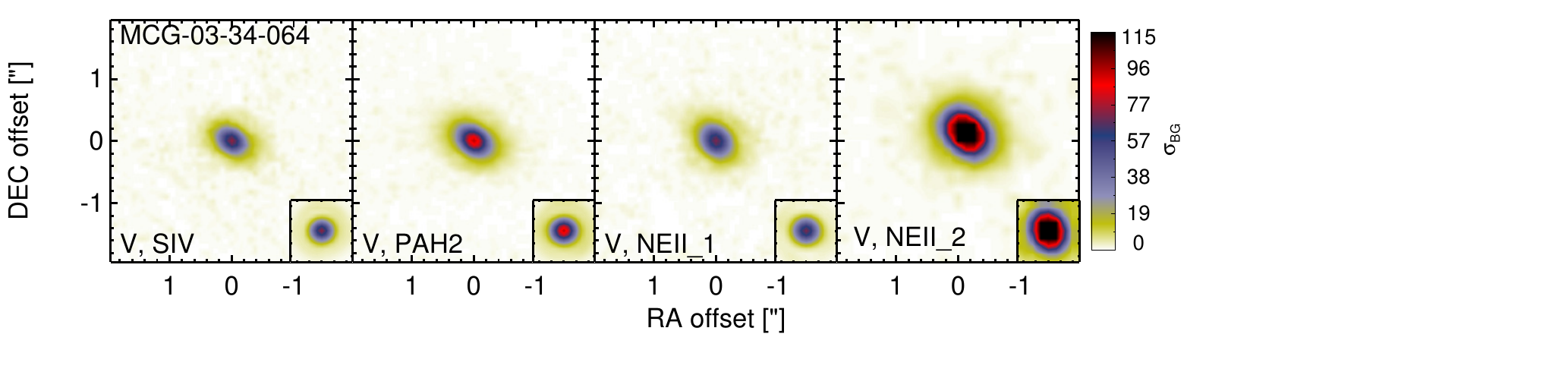} ~
\includegraphics[scale=0.9]{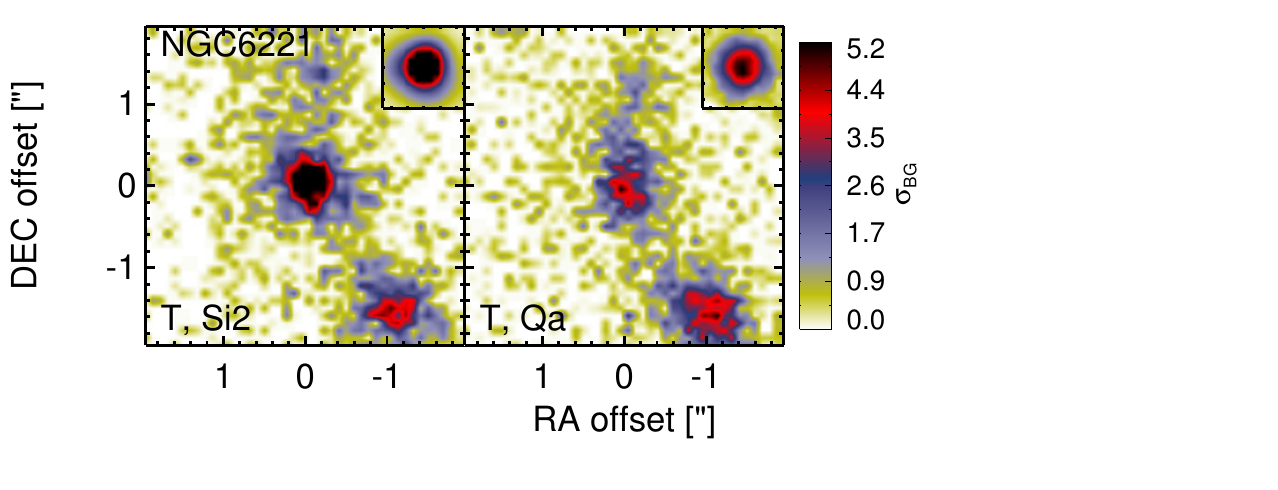} 
\caption{\label{fig1} 
Example MIR images of the inner 4 arcsec of three AGN ('M':
Michelle, 'T': T-ReCS, 'V': VISIR). 
The color scaling is linear in terms of the standard deviation of the local
background, $\sigma_\mathrm{BG}$ while white corresponds to median background
or less. 
The lower left sub-panels show the corresponding standard as an illustration of
the point spread function. 
}
\end{figure}

Out of the 249 AGN, 200 have been clearly detected. 
The majority of images show compact or point-like nuclear emission and no
host galaxy emission is detected.  
Fig.~\ref{fig1} shows examples of three AGN: NGC\,5506 has been observed and
detected with Michelle, T-ReCS and VISIR in various N- and Q-band filters. 
In all of them it appears point-like without any evidence of non-nuclear
emission. 
MCG-03-34-064 on the other hand, appears elongated in North-East direction
consistently in all images, see also \cite{Hoenig10}. 
Finally, NGC\,6221 has extended emission around the compact nucleus connecting
the latter to a bright source South-West of the nucleus, see also
\cite{RamosA11}.
This off-nuclear emission is caused by SF. 
In total, $\sim 18\%$ of the objects show consistent evidence of at least
partly resolved nuclear or circum-nuclear MIR emission.
The unresolved nuclear MIR emission measured in all images is used for
further analysis in the following.
In particular, we concentrate on the MIR emission at $\sim 12\,\mu$m measured
either directly with suitable narrow-band filters (most objects) or inferred
from adjacent filters. 
The spectral region around $12\,\mu$m is free of strong spectral features and
thus serves as the best monochromatic estimate of the nuclear MIR continuum luminosity. 

\subsection{Constraining nuclear star formation emission}
The method described in \cite{Asmus11} allows for constraining the
contribution of SF towards the nuclear $12\,\mu$m emission by using
a tight empirical correlation between the strength of the polycyclic aromatic
hydrocarbons (PAH) emission feature at $11.3\,\mu$m and the continuum MIR
emission
in starburst galaxies.
PAHs are commonly used as a SF tracer
\cite{Calzetti07}.
Therefore, it is possible to scale a SF template spectrum from
\cite{Brandl06} to the PAH flux in the individual AGN.  
This PAH flux is in turn either constrained from the lower-angular
resolution \textit{Spitzer}/IRS spectra of the AGN (if available), or a suitable HR
photometric measured covering the 11.3$\,\mu$m feature. 
This method yields an upper limit for the relative nuclear SF contribution at $12\,\mu$m, $c_<^\mathrm{SB}$.
As an example, we examine the distribution of $c_<^\mathrm{SB}$ for the observed part of the BAT AGN sample in Fig.~\ref{fig2}:
\begin{figure}[htp]
\center
\includegraphics[scale=0.9]{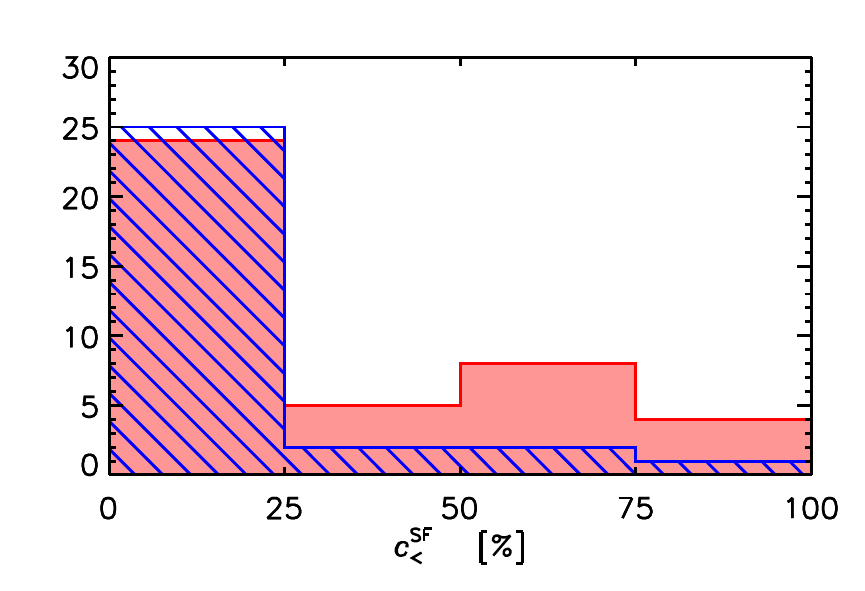} 
\caption{\label{fig2} 
Distribution of the maximum SF contribution at $12\,\mu$m for the observed 80 AGN from the BAT 9-month sample. 
Type 1(2) AGN are shown in the blue hatched (red solid) histogram.
}
\end{figure}
in the majority of BAT AGN the maximum SF contribution at sub-arcsec scales is minor.
This is in particular the case for type 1 AGN while type 2 AGN tend to exhibit higher values of $c_<^\mathrm{SB}$.

\subsection{Relation to X-ray emission}
\begin{figure}[htp]
\center
\includegraphics[scale=1.05]{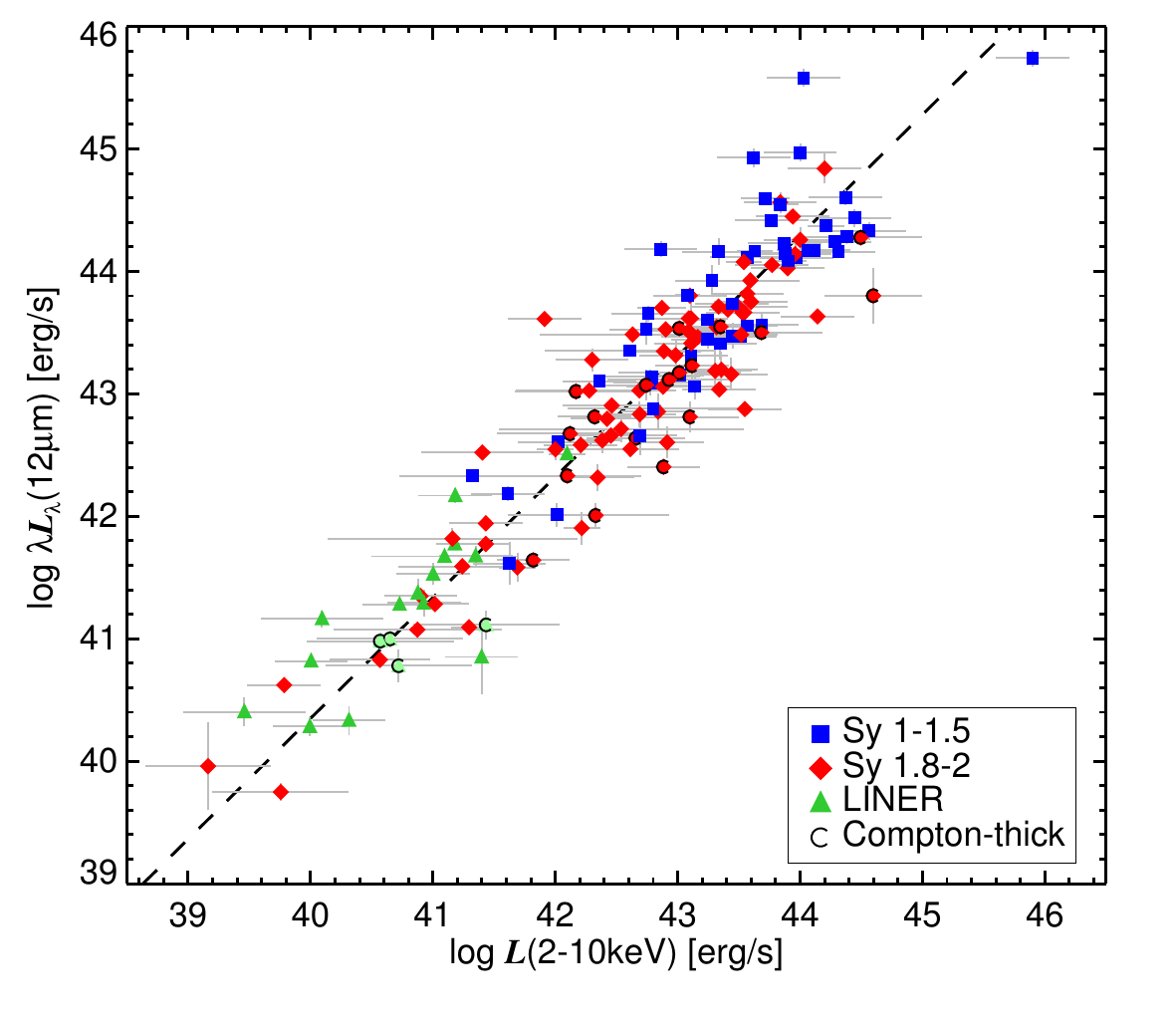}
\caption{\label{fig3} 
Relation of observed MIR and absorption-corrected X-ray luminosities for 155 detected AGN (preliminary). 
The dashed line represents the best power-law fit to all data points obtained with \texttt{linmix\_err}.
}
\end{figure}

In the following, we present preliminary results for the majority of detected AGN from the whole AGN MIR atlas.
Similar to \cite{Gandhi09,Hoenig10,Asmus11}, we have compiled
absorption-corrected 2-10\,keV X-ray luminosities, $L_\mathrm{X}$, for most of the detected AGN  from the most recent satellite missions and publications.
Details and final results for all AGN will be published in Asmus et al., in prep. b.
For 155 detected AGN of the atlas, we compare the X-ray luminosities to the observed monochromatic $12\,\mu$m luminosities, $L_\mathrm{MIR}$,   in Fig.~\ref{fig3}.
As expected from previous investigations (e.g.,\cite{Krabbe01,Gandhi09,Levenson09}), a strong correlation is found, which is best described by $\log L_\mathrm{MIR} = 0.33 \pm 0.03 + (0.99 \pm 0.03) \log L_\mathrm{X}$ (See also the contribution of Ichikawa et al.).
The observed scatter in the MIR-to-X-ray ratio is $\sim 0.42$\,dex and the intrinsic scatter is estimated to be $\sim 0.28$\,dex with \texttt{linmix\_err} \cite{Kelly07}.
The same relation is also present in flux-space and hence is not caused by distance-related effects. 

\subsubsection{Dependency on the AGN type}
\begin{figure}[htp]
\center
\includegraphics[scale=0.77]{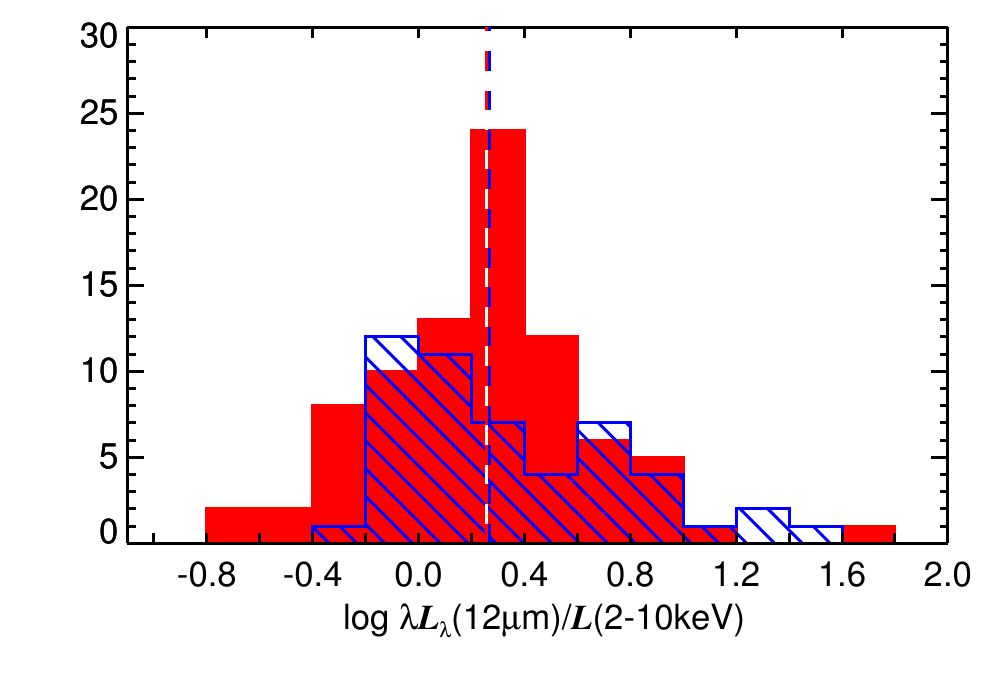}~
\includegraphics[scale=0.77]{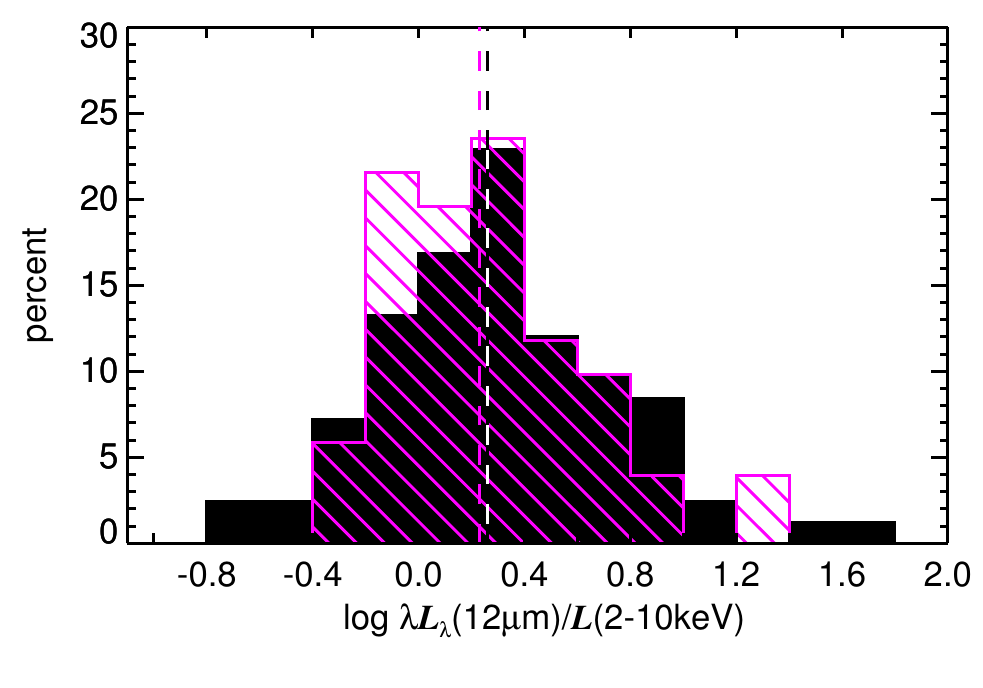}
\caption{\label{fig4} 
Distributions of the MIR--X-ray luminosity ratio with optical AGN type (left) and X-ray AGN type (right; preliminary).
Optical type 1 (2) AGN are represented by the blue hatched (red solid) histogram. 
The magenta hatched (black solid) histogram shows X-ray type 1 (2) AGN.
The dashed lines represent the corresponding median luminosity ratios.  
}
\end{figure}

Previous investigations \cite{Gandhi09,Asmus11} have found no evidence for any dependency of the MIR--X-ray correlation on the AGN type but have been limited to relatively small samples. 
Because of the linear relation between X-ray and MIR emission, such dependencies can be directly probed with the MIR--X-ray luminosity ratio. 
The distribution of this ratio is displayed for optical and also X-ray AGN types in Fig.~\ref{fig4}.
Here, we exclude the LINERs and define the X-ray AGN types as follows: 
Seyferts with a hydrogen column density, $N_\mathrm{H}$, $\le 10^{22}\,\mathrm{cm}^2$ are called unabsorbed, i.e., type~1, and objects with $N_\mathrm{H} > 10^{22}\,\mathrm{cm}^2$ absorbed, i.e., type 2. 
The distributions of the MIR--X-ray ratio of both classification schemes do not show any significant difference between absorbed and unabsorbed AGN. 
Instead, the median luminosity ratios are basically indistinguishable.

\subsubsection{Relation to X-ray obscuration}
\begin{figure}
\center
\hspace*{-0.8cm}\includegraphics[scale=0.8]{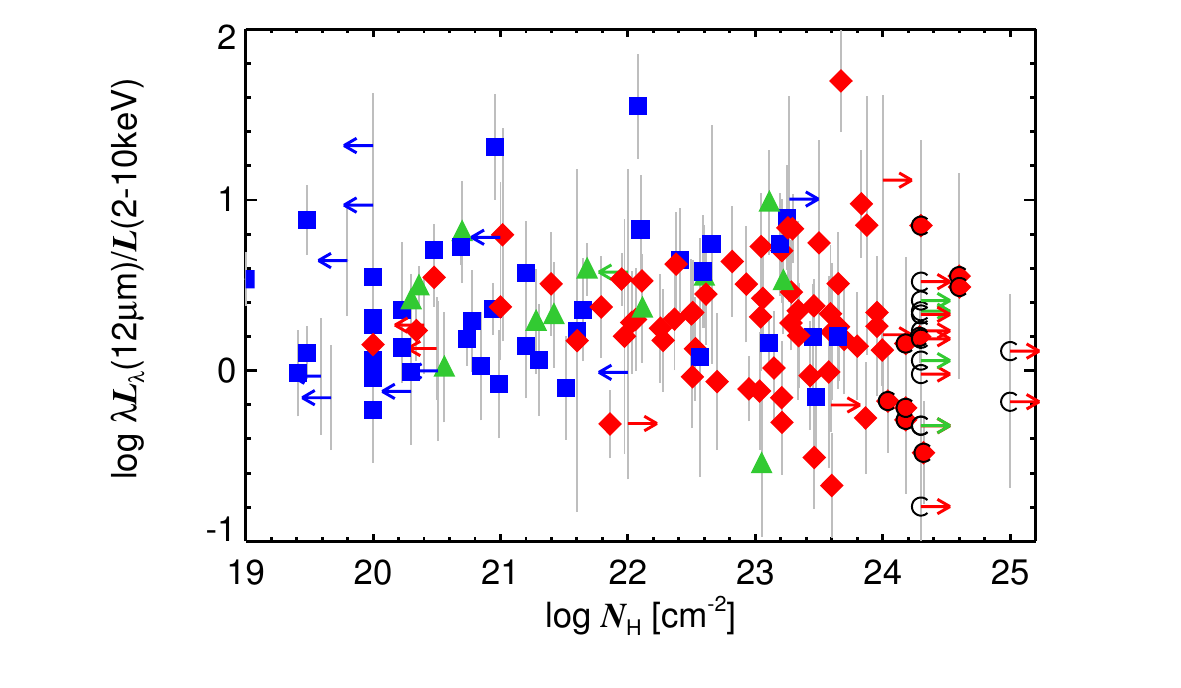}~
\hspace*{-2cm}\includegraphics[scale=0.8]{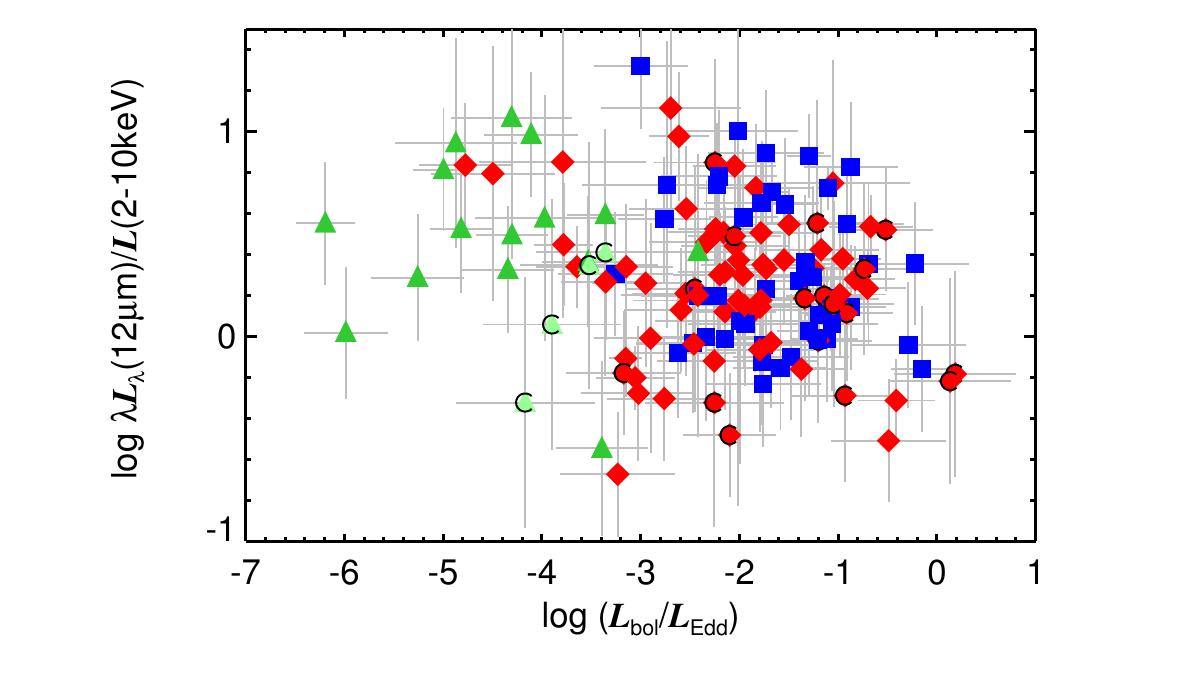}
\caption{\label{fig5} 
Relation of the MIR--X-ray luminosity ratio on the X-ray absorbing column density (left) and the Eddington ratio (right; preliminary). 
Symbols are similar to Fig.~\ref{fig3}. 
}
\end{figure}

In addition to the X-ray AGN classification, we can also directly examine the relation of the MIR--X-ray luminosity ratio with the X-ray column density, $N_\mathrm{H}$ (Fig.~\ref{fig5}). 
Optical type 1 and 2 AGN become well separated at $N_\mathrm{H} \sim 10^{22}\,\mathrm{cm}^2$ in Fig.~\ref{fig5}, indicating a good correspondence of the optical and X-ray type classification. 
Furthermore, up to column densities of at least $10^{23.3}\,\mathrm{cm}^2$, the luminosity ratio remains constant and independent of the AGN type. 
Only at highest column densities, a decreasing trend of the luminosity ratio is indicated but remains statistically insignificant. 

\subsubsection{Relation to Eddington ratio}
One of the most important fundamental parameters of AGN is the accretion rate.
It does not only determine the intrinsic object brightness but presumably also the accretion structure.
The accretion rate is commonly approximated by the Eddington ratio, $\eta = L_\mathrm{Bol} / L_\mathrm{Edd}$ where the bolometric luminosity is here simply estimated by $L_\mathrm{Bol} = 10 L_{2-10\,\mathrm{keV}}$.
The relation of the MIR--X-ray luminosity ratio with the Eddington ratio is displayed as well in Fig.~\ref{fig5} for the same 155 AGN from Fig.~\ref{fig3}. 
Type 1 and 2 AGN predominately occupy the region of $-3 < \log \eta < 0$ without any separation (median $\log L_\mathrm{MIR}/L_\mathrm{X} \sim 0.2$). 
On the other hand, most LINERs, which have $\log \eta < -3$, exhibit high MIR--X-ray ratios (median $\log L_\mathrm{MIR}/L_\mathrm{X} \sim 0.6$). 
This is also true for the type 2 AGN with the lowest Eddington ratios.
Indeed, a 1D Kolmogorov-Smirnoff test indicates a significant difference between low and high accretion rate objects with dividing threshold around $\eta \sim -3.9$. 

\section{Conclusions}
We have presented preliminary results for the AGN MIR atlas consisting of 249 local AGN with HR MIR imaging observations with four of the largest optical/infrared telescopes available today. 
In total, 200 objects have been detected and appear compact in most cases. 
By comparison with a typical starburst template we have excluded SF as the main cause of the point-like nuclear emission. 
Instead, most of the observed MIR emission appears to originate from AGN-heated dust, which still needs further verification.
We find a strong correlation of the observed nuclear MIR emission with the absorption-corrected X-ray emission. 
Such a correlation can be approximately understood by a link between the MIR and X-ray emission via reprocessing of primary radiation from the accretion disk, which is strongest in the ultraviolet (UV). 
The latter is reprocessed on the one hand in the inner most regions around the SMBH, where the UV photons are up-scattered from hot corona electrons into the X-ray regime.
And on the other hand, much further out in the torus region, the UV photons are absorbed by dust and reemitted as thermal radiation in the MIR. 
However, more detailed investigations of the MIR--X-ray luminosity ratio do not show any evidence for a dependency on AGN luminosity or type. 
Therefore, scenarios like lower covering factors for higher luminosities or type 1 AGN are not supported (see also contribution by Lawrence et al.).  
In addition, no evidence for a dependency of the MIR--X-ray ratio with nuclear obscuration as expected from torus models is found. 
Nevertheless, it is possible that MIR--X-ray luminosity is not dominated by one but by several effects, in which case the apparent constancy of this ratio would actually be caused by their combination.
Clearly, further investigations are needed.
Finally, there seems to be a dependency on the accretion rate, namely a possible structural change at lowest accretion rates, i.e. the occurrence of a strong jet component.
The latter is known to produce copious amounts of synchrotron emission also in the MIR emission which easily dominate the whole spectrum (see e.g. contribution by Perlman et al.).

%
\small  
%
%

%

%
\end{document}